\documentclass{article}

\usepackage{PRIMEarxiv}
\usepackage{mathrsfs}
\usepackage[utf8]{inputenc} 
\usepackage[T1]{fontenc}    
\usepackage{hyperref}       
\usepackage{url}            
\usepackage{booktabs}       
\usepackage{amsfonts}       
\usepackage{amsmath,amssymb}
\usepackage{nicefrac}       
\usepackage{microtype}      
\usepackage{lipsum}
\usepackage{fancyhdr}       
\usepackage{graphicx}       
\graphicspath{{media/}}     
\usepackage[]{natbib}
\usepackage{bm}
\usepackage{multirow}
\usepackage{float} 
\usepackage{algorithm}
\usepackage{algorithmic}
\title{Estimating causal effects of continuous-time dynamic treatments with unmeasured confounders}

\author{
  Haiyan Zhu$^1$, Yingchun Zhou$^*$\\
  Key Laboratory of Advanced Theory and Application in Statistics and Data Science -\\ MOE, School of Statistics, East China Normal University,\\
   200062, Shanghai, China.\\
  \texttt{$^*$yczhou@stat.ecnu.edu.cn} \\
}

\begin{document}
\maketitle

\begin{abstract}
Modern medical research demands specialized causal inference methods evaluating complex continuous-time dynamic treatment regimens using observational data. For instance, obtaining the causal effects of intravenous administration, a continuous process involving dynamic adjustments of the treatment dose, can guide clinicians on drug use. However, the existing causal inference frameworks in longitudinal studies typically assume that time advances in discrete time steps. Therefore, this paper proposes a new methodology to estimate the causal effects of continuous-time dynamic treatments in the presence of unmeasured confounding. Unmeasured confounding is incorporated into estimating continuous-time Marginal Structural Models from a Bayesian perspective. Simulation demonstrates that compared to existing methods, the proposed approach can provide approximately unbiased estimates for target causal parameters across three degrees of confounding. The proposed method is applied to analyze the causal relationship between the intravenous oxytocin administration process and postpartum hemorrhage, leading to meaningful results that may guide clinicians in using oxytocin.
\end{abstract}


\section{Introduction}\label{sec1}
In clinical practice, dynamic treatment regimens are commonly used \cite{cain2010start},  and there has been an increasing focus on dynamic treatments that change continuously over time (continuous-time dynamic treatments) \cite{zhang2011causal,ying2022causal}. Therefore, modern medical studies call for advanced methodologies to evaluate the causal effects of complex continuous-time dynamic treatments. This work is motivated by clinical studies on the effects of the oxytocin administration process on postpartum hemorrhage (PPH) \cite{zhu2024oxytocin}. PPH is a significant factor of maternal morbidity and mortality worldwide \cite{say2014global}. Widespread concerns have been raised about whether prolonged use of oxytocin might be linked to a higher risk of PPH \cite{zhu2024oxytocin}. Therefore, estimating the causal effects of the oxytocin administration process on postpartum hemorrhage is crucial in guiding the use of oxytocin.

\begin{figure*}
	\centerline{\includegraphics[width=350pt,height=15pc]{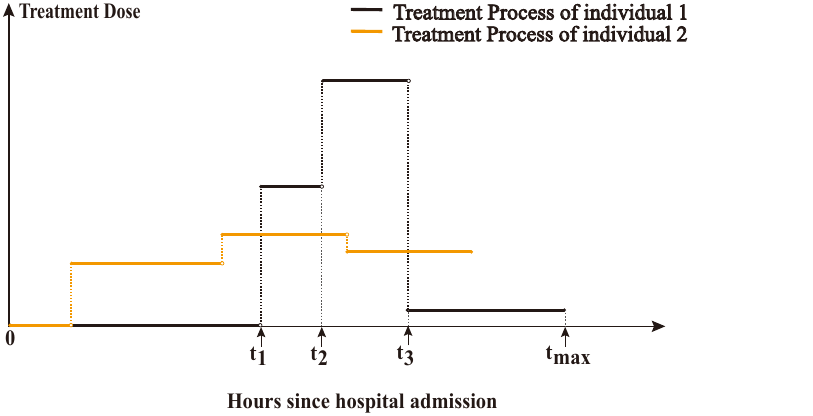}}
	\caption{Illustration of oxytocin administration processes for two individuals.}
	\label{fig:Fig1}
\end{figure*}

The complex nature of the oxytocin administration process, owing to different choices of timing and dosage, poses three major challenges for current causal inference frameworks. First, treatment status over time may depend on evolving patient- and disease-specific covariates. Second, treatments are measured continuously over time. Third, unobserved confounding variables may exist at each time point, which is common in observational data. Figure \ref{fig:Fig1} illustrates the oxytocin administration processes for two individuals, which may help to understand continuous-time dynamic treatments better. It shows that the number and timing of treatment adjustments vary across individuals. This fundamentally contrasts with traditional longitudinal treatments with fixed observation time points. The individual, represented by the black line segment in Figure \ref{fig:Fig1}, experiences three dose adjustments (at time $t_1,t_2$ and $t_3$), and the delivery is completed at time $t_{\max}$. Changes in dose depend on the subject's history information of baseline (e.g., age and weight) and time-varying covariates (e.g., cervical dilation).

Extensive theories around causal inference for discrete-time longitudinal data have been developed \cite{robins1997causal, robins2000marginal,robins2000marginalversus,Hernan_robins_2020,Hernán_2001_06}. Among these methods, Marginal Structural Models (MSMs, \cite{robins2000marginal, Hernán_2001_06}) coupled with weighting offer several advantages. They help control for measured confounding and avoid the need to model potentially high-dimensional intermediate variables explicitly. Saarela et al. \cite{saarela2015bayesian} introduced a Bayesian approach for estimating parameters in MSMs within a discrete-time framework. Their methodology can handle latent individual-level "frailty" variables, which influence both the outcome and intermediate variables but are assumed to be independent of treatment assignments.

The recent causal inference literature has seen several attempts to develop identification frameworks for continuous-time longitudinal data. Lok \cite{lok2008statistical} provides a conceptual framework and formalization for structural nested models in continuous time without practical implementation. Rytgaard, Gerds, and van der Laan \cite{rytgaard2022continuous} study the generalization of the targeted minimum loss-based estimation (TMLE) framework to the estimate of effects of time-varying interventions in settings where both interventions, covariates, and the outcome can happen at subject-specific time-points on an arbitrarily fine timescale.  Hu and Hogan \cite{hu2019causal}, Ryalen et al. \cite{ryalen2020causal}, and Hu et al. \cite{hu2023estimating} have demonstrated the effectiveness of continuous-time marginal structural models in addressing time-varying confounding and providing consistent causal effect estimators. However, the studies mentioned above require the assumption of no unmeasured confounding.  

A new Bayesian framework is proposed in this paper to estimate the causal effects of continuous-time dynamic treatments in the presence of unmeasured confounding. We carefully define the likelihood and construct the outcome model. Specifically, we establish the likelihood within the continuous-time framework that aligns with the data structure detailed in this paper. As for the outcome model, we construct it  with the hope of simultaneously characterizing the effects of time and treatment dose. 

This work makes two major contributions to the literature on causal inference. First, we develop a Bayesian framework capable of estimating the causal effects of continuous-time dynamic treatments, which is rarely studied. Second, the proposed framework can handle unmeasured confounding with continuous-time dynamic treatments, which is an issue that has not been addressed so far in the literature on causal inference.

The article is organized as follows. Section \ref{sec2} introduces the notation and setup. Section \ref{sec3} describes the process of constructing the framework. First, it outlines the form of the outcome model and specifies the target causal parameters. Then, it builds the likelihood function and finally derives the posterior distributions of target causal parameters. Simulation in Section \ref{sec4} compares the proposed approach with existing methods across three levels of confounding. Section \ref{sec5} analyzes the causal relationship between the intravenous oxytocin administration process and postpartum hemorrhage.  A discussion is provided in Section \ref{sec6}.

\section{NOTATION AND SET UP}\label{sec2}
Consider a longitudinal observational study setting involving the individuals $i = 1,...,n$, with treatment decisions continuously and dynamically carried out at some fixed and finite time interval $\mathscr{T}=[0,t_R]$. Denote $T_{max}$ as the time point when the treatment ends, with $\{N^{T_{max}}(t): 0\le t\le t_R\}$ as its associated zero-one counting process \cite{andersen2012statistical}. $N^{T_{max}}(t)=1$ means the treatment has already ended at time $t$ or earlier, and $N^{T_{max}}(t)=0$ otherwise. Under the assumption $0< T_{max}\le t_R$, we have $N^{T_{max}}(0)=0$ and $N^{T_{max}}(t)=1$ for any $T_{max}\le t\le t_R$. Let $\lambda^{T_{max}}(t)$ be the corresponding intensity function.

Denote $\{A(t) \in \mathcal{A}: t \in \mathscr{T}\}$ as the treatment process, $\bar{A}(t)=\{A(s):0\le s\le t\}$ as the treatment history up to and including time $t$, and $\bar{A}$ as the  abbreviation of $\bar{A}(t_R)$. $1\{\cdot\}$ denotes the indicator function, and $t-$ represents the left-hand limit at time $t$. Let $\Delta_A(t)=1\{A(t)-A(t-)\ne 0\}$ describe whether a change in treatment dose occurs at time $t$, and $N^A(t)$ be the counting process that records the number of changes in the treatment status (i.e., the count of $\Delta_A(t)$ being equal to one) up to and including time $t$. $\lambda^A(t)$ is the intensity function corresponding to $N^A(t)$. Define $A(t)$ to be zero for all $t > T_{\max}$. $\mathcal{A}$ is the set of all possible values of $A(t),t\in \mathscr{T}$ and $\mathcal{\bar{A}}$ is the set of all possible values of $\bar{A}$.

In addition, let $Y\in \mathbb{R}$ denote the outcome measured at the end of the study, and $Y_{(\bar{a},t_{max})}\in \mathbb{R}$ denote the counterfactual outcome if the treatment process is set as $\bar{a}$ and treatment ends at time $t_{max}$. Suppose each individual has a $p_Z$ dimensional set of baseline covariates $Z$ (such as age, height, and weight) and a $p_L$ dimensional discrete-time varying covariate process $\bar{L}=\{L(t)\in\mathcal{L}: 0\le t\le T_{max}\}$. For simplicity of exposition, this paper assumes $p_L = 1$. Assume $L(t)$ changes status at a finite discrete-time set $D_L=\{0=t^L_0< t^L_1<t^L_2<\cdots<t^L_{k_L}<t_R=t^L_{k_L+1}\}$. This assumption imposes no serious practical limitations because the flexibility can be adjusted by $k_L$. For $t\in (t^L_k,t^L_{k+1}) $, set $L(t)=L(t^L_k), k=0,...,k_L$.
Let $\bar{L}(t)=\{L(s)\in\mathcal{L}: 0\le s\le t\}$ denote the observed history information of $\bar{L}$ up to and including time $t$, where $\mathcal{L}$ is the set of all possible values of $L(t),t\in D_L$.


Assume unmeasured confounders $U$ with dimension $p_U$ influence both $\{A(t),t\in [0, T_{max}]\}$ and $Y$. The shorthand notation $v_i=(y_i,t_{max,i},z_i,\bar{a}_i,\bar{l}_i)$ and $\tilde{v}_i=(y_i,t_{max,i},z_i,u_i,\bar{a}_i$,
$\bar{l}_i), i=1,...,n$ represent the observed and complete variables respectively. Denote $v$ and $\tilde{v}$ without subscript as the corresponding vectors for $n$ observations.  

\section{METHODOLOGY}\label{sec3}
Denote the notation $\mathcal{J_O}$ as the  {\itshape{observational}} world where the treatment assignment can depend on covariates, $\mathcal{J_E}$ as the {\itshape{experimental}} world where the treatment assignment is not influenced by covariates (i.e., causal inference may be performed). Causal inferences are possible if the continuous-time dynamic treatment effects under $\mathcal{J_E}$ can be estimated based on the data observed under $\mathcal{J_O}$. The following sections establish a connection between two worlds through a Bayes decision rule, with the aim of estimating causal parameters.

\subsection{Assumptions}
The basic assumptions that will be used throughout the paper are as follows, which are extensions of the usual causal assumptions \citep{robins1999association, robins2000marginal}:

$(A1)$ (Continuous-Time Consistency) For any treatment regimes $(\bar{a},t_{max}), \bar{a}\in\mathcal{\bar{A}}, 0<t_{max}\le t_R$, $Y=Y_{(\bar{a},t_{max})}$, almost surely.

$(A2)$ (Continuous-Time Positivity, \cite{rytgaard2022continuous}) 
Denote the distribution of the observed data as $P$ and decompose $P$ into the interventional part ($G$) and the non-interventional part ($Q$) (i.e. $dP=dP_{Q,G}$). Assume absolute continuity of $P_{Q,G^*}$ with respect to $P_{Q,G}$ (i.e., $P_{Q,G^*}<<P_{Q,G}$), where $G^*$ is a user-specified treatment regime. This implies existence of the Radon-Nikodym derivative $dP_{Q,G^*} /dP_{Q,G}$.

$(A2')$ (Bayesian Continuous-Time Positivity) The ratio $w^*_i = p(v^*_i\mid v,\mathcal{J_E})/p(v^*_i\mid v,\mathcal{J_O})$ is well-defined, where $v_i^*=(y^*_i,t^*_{max,i},z^*_i,\bar{a}^*_i,\bar{l}^*_i)$ is from the super-population (or generating mechanism) characterized by $p(v\mid \mathcal{J_E})$.

$(A3)$ (Continuous-Time Latent Conditional Exchangeability) Assume $\forall t\in \mathscr{T}$,\\
(1) $\lambda^A(t\mid \bar{A}(t-),\bar{L}(t),U,Z,Y_{(\bar{a},t_{max})})=\lambda^A(t\mid \bar{A}(t-),\bar{L}(t),U,Z)$;\\
(2) $p(a^* \mid \Delta_A(t)=1, \bar{A}(t-),\bar{L}(t),U,Z,Y_{(\bar{a},t_{max})})=p(a^* \mid \Delta_A(t)=1, \bar{A}(t-),\bar{L}(t),U,Z)$, where $a^*=A(t)\in \mathcal{A}$ and $a^*\ne A(t-)$;\\
(3) $\lambda^{T_{max}}(t\mid \bar{A}(t-),\bar{L}(t-),Z,Y_{(\bar{a},t_{max})})=\lambda^{T_{max}}(t\mid \bar{A}(t-),\bar{L}(t-),Z)$.

Assumption  $(A1)$ establishes a connection between the observed outcome and the potential outcome through the treatment regimen actually received. It says that if an individual receives the treatment regime $(\bar{a},t_{max})$, then his/her observed outcome $Y$ is the same as the potential outcome $Y_{(\bar{a},t_{max})}$. Assumption $(A2)$ implies that there are enough data to make causal inference under the target distribution $P_{Q,G^*}$ \cite{ying2022causal}. The variables of the interventional part in this paper consist of $\bar{A}$ and $T_{max}$. Assumption $(A2')$ is the Bayesian counterpart of Assumption $(A2)$. The meaning of Assumption $(A2')$ will be further explained in the subsequent sections. In Assumption $(A3)$, unmeasured confounders $U$ are assumed to have no influence on $T_{max}$ for the sake of simplicity in exposition. Our method can be easily extended to the scenario that $T_{max}$ is also influenced by $U$. In addition, conditions (1) and (2) in Assumption $(A3)$ characterize the "Latent Conditional Exchangeability" for $A(t)$ from the perspective of the intensity function and the distribution at jump points.

\subsection{MSMs and target causal parameters}
\label{sec3.1}
In this section, the potential outcome model is defined, and the target causal parameters of interest are specified.

Marginal structural models are formulated as marginal distributions of potential outcomes, which are functionally dependent on hypothetical treatment interventions. We consider MSMs of the form:

\begin{equation}
	Y_{(\bar{a},t_{max})}=\eta_1+\eta_2\int_0^{t_{max}}a(t)\exp(-\eta_3\frac{|t-t_{max}|}{t_{max}})dt+\epsilon, \tag{3.1} \label{equa_3_1}
\end{equation}
where $\eta_1,\eta_2,\eta_3$ are coefficients ($\eta_3\ge0$), $\epsilon\sim N(0,\sigma^2)$ is the error. 
In the potential outcome model (\ref{equa_3_1}), both the time and dose of intervention can influence the potential outcome. Specifically, the treatment dose effect is denoted by the coefficient $\eta_2$, and the time effect is controlled by the coefficient $\eta_3$. 

One of the main reasons for structuring the model in this way is that the assumption of exponentially decaying time effects has been widely adopted. For example, Hawkes processes \cite{hawkes1971point} have been successfully applied across various domains (e.g., \cite{ogata1988statistical, tucker2019handling, kalair2021non, rambaldi2017role}). A popular choice of influence kernel, which determines the spread of influence over time, is the exponential kernel, known for its strong performance in numerous applications (e.g., \cite{shelton2018hawkes, deutsch2022estimating}). Similarly, to model the accumulated effect of certain drugs, exponential decay is often assumed \cite{hua2022personalized}. In addition, Pors Nielsen et al. \cite{nielsen1994magnitude} employed an exponential model to describe the change in bone mineral density observed in response to estrogen in postmenopausal women.

The causal parameter vector $\eta=(\eta_1,\eta_2,\eta_3)^T$ can be estimated using its observed counterpart $p(y_i\mid t_{max,i},\bar{a}_i,\eta)$ under a data generating mechanism without confounding. When confounding exists, Section \ref{sec3.3} shows that under Assumptions $(A1)$, $(A2)$ and $(A3)$, $\eta$ may be estimated by maximizing the IPT-weighted pseudo-likelihood function.
\subsection{Likelihood construction}
\label{sec3.3}
The critical aspect of conducting Bayesian analysis is defining the likelihood function for the data. This section describes the process of constructing the likelihood function and the continuous-time marginal structural models.

The challenge in constructing the likelihood function lies in describing the treatment process and the zero-one counting process related to $T_{max}$. Let $\Lambda^X$ denote the cumulative intensity that characterizes the compensator of $N^X,X\in \{A,T_{max}\}$. Denote $\mathcal{H}^A_{t},\mathcal{H}^{T_{max}}_{t}$ as the history information that influence $A(t)$ and $d\Lambda^{T_{max}}$ at time $t$, respectively. Assume the realizations of all processes are càdlàg functions on the corresponding intervals, and there are no events at the start time. Heuristically, we have that $$P(N^X(dt)=1\mid \mathcal{H}^X_{t})=E[N^X(dt)\mid\mathcal{H}^X_{t}]=d\Lambda^X(t\mid\mathcal{H}^X_{t})=\lambda^X(t\mid\mathcal{H}^X_{t})dt,$$ where $\lambda^X$ is the intensity function and the increment $N^X(dt)$ is non-zero and equal to 1 iff there is a jump of $N^X$ in the infinitesimal interval $[t,t+dt)$ \cite{andersen2012statistical}. 

Characterizing the treatment process requires not only describing the number of times of dose adjustments but also the specific values. So we use a mixture distribution to model $A(t)$ at any infinitesimal interval $[t,t+dt)$:
$$[\lambda^A(t\mid\mathcal{H}^A_{t})dt\times p(A(t)\mid\mathcal{H}^A_{t})]^{N^A(dt)=1}[1-\lambda^A(t\mid\mathcal{H}^A_{t})dt]^{N^A(dt)=0}.$$
Going from one infinitesimal interval to the next, one can derive the limit expression with the first quantity being equal to the finite product over the jump times and the second quantity being the survival function for the treatment adjustment:
$$\prod\limits_{t_j:\Delta_{A}(t_j)=1}[\lambda^A(t_j\mid\mathcal{H}^A_{t_j})p(A(t_j)\mid\mathcal{H}^A_{t_j})]\exp\{-\int_0^{T_{max}}\lambda^A(t\mid\mathcal{H}^A_t)dt\}.$$

In our setting, $N^{T_{max}}(t_R)$ can take two values: (i) $N^{T_{max}}(t_R)=1$, which means the treatment is as expected terminated between $(0,t_R]$. In this case, $T_{max}=t_{max}\in(0,t_R]$. (ii) $N^{T_{max}}(t_R)=0$, which means the treatment does not end within  $(0,t_R]$. In this case, we denote $T_{max}=t_{max}=t_R$. Corresponding to the two cases, the part of $T_{max}$ in the likelihood can be written as
$$[\lambda^{T_{max}}(t_{max}\mid\mathcal{H}^{T_{max}}_{t_{max}})]^{1\{N^{T_{max}}(t_R)=1\}}\exp\{-\Lambda ^{T_{max}}\},$$ 
where $\Lambda^{T_{\text{max}}} =\int_{0}^{t_{\text{max}}} \lambda^{T_{\text{max}}}(t\mid\mathcal{H}^{T_{\text{max}}}_{t}) \, dt$.

Let $\theta=(\theta_{T_{max}},\theta_{A},\theta_Z,\theta_L,\theta_U,\theta_Y)$ represent model parameters under $\mathcal{J_O}$ and $\alpha=(\alpha_{T_{max}},\alpha_{A})$ denote parameters of the interventional part under $\mathcal{J_E}$. Further, denote $\alpha_A=(\alpha_{A1},\alpha_{A2})$ as a partitioning of $\alpha_A$, where $\alpha_{A1}$ and $\alpha_{A2}$ specify $\lambda^A(t\mid\mathcal{H}^A_t)$ and $p(A(t)\mid\mathcal{H}^A_t)$, respectively. Similarly, a partitioning for $\theta_A$, $\theta_A=(\theta_{A1},\theta_{A2})$, is also defined. It is worth noting that the model parameters of the non-intervention part are assumed to be the same across  $\mathcal{J_O}$ and $\mathcal{J_E}$. To facilitate concise expression in subsequent formulas, we define the following notations, which are components of the likelihoods under $\mathcal{J_O}$ and $\mathcal{J_E}$:

\begin{itemize}
	\item $p^{\mathcal{J_E}}_{A,i}=\prod\limits_{t_{j}:\Delta_{a_i}(t_{j})=1}[\lambda^A(t_{j}\mid\mathcal{H}^{A,\mathcal{J_E}}_{it_j},{\alpha}_{A1})p(a_i(t_{j})\mid\mathcal{H}^{A,\mathcal{J_E}}_{it_j},{\alpha}_{A2})]\exp\{-\Lambda^{A,\mathcal{J_E}}_i\}$\\
	$p^{\mathcal{J_O}}_{A,i}=\prod\limits_{t_{j}:\Delta_{a_i}(t_{j})=1}[\lambda^A(t_{j}\mid\mathcal{H}^{A,\mathcal{J_O}}_{it_j},{\theta}_{A1})p(a_i(t_j)\mid\mathcal{H}^{A,\mathcal{J_O}}_{it_j},{\theta}_{A2})]\exp\{-\Lambda^{A,\mathcal{J_O}}_i\}$
	\item $p^{\mathcal{J_E}}_{T_{max},i}=[\lambda^{T_{max}}(t_{max,i}\mid\mathcal{H}^{T_{max},\mathcal{J_E}}_{it_{max,i}},\alpha_{T_{max}})]^{1\{N^{T_{max,i}}(t_R)=1\}}\exp\{-\Lambda^{T_{max},\mathcal{J_E}}_i\}$\\
	$p^{\mathcal{J_O}}_{T_{max},i}=[\lambda^{T_{max}}(t_{max,i}\mid\mathcal{H}^{T_{max},\mathcal{J_O}}_{it_{max,i}},\theta_{T_{max}})]^{1\{N^{T_{max,i}}(t_R)=1\}}\exp\{-\Lambda_i ^{T_{max},\mathcal{J_O}}\}$
	\item $p_{Y,i}=p(y_i\mid t_{max,i},\bar{a}_i,\bar{l}_i,z_i,u_i,\theta_Y)$
	\item $p_{Z,i}=p(z_i\mid {\bf{\theta}}_Z)$
	\item $p_{L,i}=\prod\limits_{t_{j}\in D_L\cap [0,t_{max,i}]}p(l_i(t_{j})\mid\mathcal{H}^L_{it_{j}},\theta_L)$
	\item $p_{U,i}=p(u_i\mid{\bf{\theta}}_U)$
\end{itemize}
where $\Lambda^{X,\mathcal{J_O}}_i$ and $\Lambda^{X,\mathcal{J_E}}_i$ denote the cumulative intensity of individual $i$ under $\mathcal{J_O}$ and $\mathcal{J_E}$, respectively, $X\in \{A,T_{max}\}$. $\mathcal{H}^{A,\mathcal{J_O}}_{it}=\{\bar{a}_i(t-),\bar{l}_i(t),u_i,z_i,t\le t_{max,i}\}$, $\mathcal{H}^{A,\mathcal{J_E}}_{it}=\{\bar{a}_i(t-), t\le t_{max,i}\}$, $\mathcal{H}^{T_{max},\mathcal{J_O}}_{it}=\{\bar{a}_i(t-),\bar{l}_i(t-),z_i\}$,  $\mathcal{H}^{T_{max},\mathcal{J_E}}_{it}=\{\bar{a}_i(t-)\}$ and
$\mathcal{H}^{L}_{it}=\{\bar{a}_i(t-),\bar{l}_i(t-),z_i,t\le t_{max,i}\}$ are the history information sets influencing corresponding variables at time $t$. The rationale for configuring these sets in this manner is that $L(t)$ is assumed to be independent of $U$ and under $\mathcal{J_E}$, the variables of the interventional part ($\bar{A}$ and $T_{max}$) are unaffected by confounders.

Finally, the likelihood of $\tilde{v}$, the vector of complete variables for $n$ observations, under $\mathcal{J_O}$ and $\mathcal{J_E}$ can be expressed as (\ref{eq_3_2}) and (\ref{eq_3_3}),  respectively:
\begin{equation}
	\prod\limits_{i=1}^n\left\{p_{Y,i}p_{U,i}p_{Z,i}p_{L,i}p^{\mathcal{J_O}}_{A,i}p^{\mathcal{J_O}}_{T_{max},i}\right\} \tag{3.2} 
	\label{eq_3_2}
\end{equation}

\begin{equation}
	\prod\limits_{i=1}^n\left\{p_{Y,i}p_{U,i}p_{Z,i}p_{L,i}p^{\mathcal{J_E}}_{A,i}p^{\mathcal{J_E}}_{T_{max},i}\right\} \tag{3.3} 
	\label{eq_3_3}
\end{equation}
The likelihood is decomposed into the product of variable-specific terms, and the difference between the two mechanisms lies in whether the intervention variables are affected by confounding.

Under Assumptions $(A1)$, $(A2)$ and $(A3)$, the causal parameter $\eta$ may be estimated by maximizing the IPT-weighted pseudo-likelihood function (cf. Hu et al. \cite{hu2023estimating}): $$q(\eta;\tilde{v},\alpha_A,\alpha_{T_{max}},\theta_{A},\theta_{T_{max}})=\prod\limits_{i=1}^np(y_i\mid t_{max,i},\bar{a}_i,\eta)^{w_i},$$
where
$w_i=\frac{p^{\mathcal{J_E}}_{T_{max},i}p^{\mathcal{J_E}}_{A,i}}{p^{\mathcal{J_O}}_{T_{max},i}p^{\mathcal{J_O}}_{A,i}}$ defines "stabilized" weights. The effect of weighting is to construct a pseudo-population in which the interventional part is not influenced by confounding \cite{robins2000marginal}. The weights reflect the extent to which the observed population resembles the target pseudo-population.

However, the weights cannot be calculated using the observed data $v_i=(y_i,t_{max,i},z_i,\bar{a}_i,\bar{l}_i),$
$i=1,...,n$ due to the unmeasured confounders $U$. The next section will solve this problem from a Bayesian perspective.
\subsection{Continuous-time IPT weighting derived through a Bayes decision rule}
\label{sec3.5}

In order to motivate Bayesian inference for target causal parameters via a utility maximization framework, the de Finetti representations \cite{bernardo2009bayesian} under $\mathcal{J_O}$ and $\mathcal{J_E}$ are first deduced in this section. It is followed by a derivation of weights that act as importance sampling weights in predicting the outcome in a hypothetical population without covariate imbalances, that is, $\mathcal{J_E}$.

Assume the complete variables $\tilde{v}_1,...,\tilde{v}_n$ are part of an \textit{infinite exchangeable} \cite{bernardo2009bayesian} sequence, one can deduce the de Finetti representation for the joint distribution of a random sample $v_1,...,v_n$ from a super-population characterized as $\mathcal{J_O}$:
$$
\begin{aligned}
	p(v\mid\mathcal{J_O})&=\int_{\theta}\prod\limits_{i=1}^n\left\{\left[\int_{u_i}p_{Y,i}\times p^{\mathcal{J_O}}_{A,i}\times p_{U,i}du_i\right]\times p_{L,i}\times p_{Z,i}\times p^{\mathcal{J_O}}_{T_{max},i}\right\}p(\theta)d\theta.
\end{aligned}
$$ 
The missing $U$ is marginalized in $p(v\mid\mathcal{J_O})$. For binary cases, the integral $\int_{u_i}$ becomes a summation $\sum_{u_i}$ and the
prior of $\theta_U$, denoted as $p(\theta_U)$, can be taken as $\text{Uniform}(0,1)$. For continuous distributions, the computational burden is relatively heavy, and additional restrictions on $U$ or prior information about $\theta_U$ may be needed to improve MCMC performance. For example, if we have an external dataset and obtain the maximum likelihood estimate $\hat{\theta}_{U, MLE}$ and variance-covariance matrix $\hat{\Sigma}_{\theta, MLE}$ in the external data, then $N(\hat{\theta}_{U, MLE},\hat{\Sigma}_{\theta, MLE})$ can be a sensible choice for $p(\theta_U)$ (cf. \cite{comment2022bayesian}).

Assumption $(A3)$ allows us to model probabilities of the interventional part in $p(v\mid\mathcal{J_O})$ with observed covariates and unobserved $U$. Assumption $(A1)$ enables us to infer counterfactual outcomes based on observational data. 

Similarly, the de Finetti representation for the joint distribution of a random sample $v_1,...,v_n$ from a super-population characterized as $\mathcal{J_E}$ can be given as 
$$
\begin{aligned}
	p(v\mid\mathcal{J_E})&=\int_{\theta^-,\alpha}\prod\limits_{i=1}^n\left\{\left[\int_{u_i}p_{Y,i}\times p_{U,i}du_i\right]\times p_{L,i}\times p_{Z,i}\times p^{\mathcal{J_E}}_{A,i}\times p^{\mathcal{J_E}}_{T_{max},i}\right\}p(\theta^{-},\alpha)d\theta^-d\alpha,
\end{aligned}
$$
where $\theta^-=(\theta_Y,\theta_U,\theta_Z,\theta_L).$ The prior distributions of the parameters  related to the two representations (with finite dimensions for simplicity) are assumed to be absolutely continuous with respect to the Lebesgue measure with density $p(\theta)$ and $p(\theta^{-},\alpha)$.

As in Saarela et al. \cite{saarela2015bayesian}, in order to make causal inferences, one needs to hypothesize generating predictions $v_i^*=(y^*_i,t^*_{max,i}, z^*_i, \bar{a}^*_i, \bar{l}^*_i)$ from the super-population where the data generating mechanism is characterized by $p(v\mid\mathcal{J_E})$, based on the observed sample $v$ of size $n$ from $p(v\mid\mathcal{J_O})$. Based on $v_i^*$, the causal parameters can be estimated with no confounding. Let $H(\cdot)$ be a utility function relevant to the estimation problem. Then
$$
\begin{aligned}
	E[H(v_i^*)\mid v,\mathcal{J_E}]&=\int_{v_i^*}H(v_i^*)p(v_i^*\mid v,\mathcal{J_E})dv_i^* \\
	&=\int_{v_i^*}H(v_i^*)\frac{p(v_i^*\mid v,\mathcal{J_E})}{p(v_i^*\mid v,\mathcal{J_O})}p(v_i^*\mid v,\mathcal{J_O})dv_i^*\\
	&=\int_{v_i^*}w_i^*H(v_i^*)p_n(v_i^*)dv_i^*,
\end{aligned}
$$
where $p_n$ is taken to be a nonparametric posterior predictive density \cite{walker2010}, and
$w^*_i = p(v^*_i\mid v,\mathcal{J_E})/p(v^*_i\mid v,\mathcal{J_O})$. Assumption $(A2')$ ensures that the ratio $p(v^*_i\mid v,\mathcal{J_E})/p(v^*_i\mid v, \mathcal{J_O})$ is well-defined.

Choosing the utility function $H(v_i^*;\eta)=\log p(y_i^*\mid t^*_{max,i},\bar{a}^*_i,\eta)\equiv l(y_i^*\mid t^*_{max,i},$
$\bar{a}^*_i,\eta)$,  and adopting the Bayesian bootstrap strategy $p_n(v^*_i ) =\sum\limits_{k=1}^n\pi_k\delta_{v_k}(v_i^*)$ \cite{walker2010}, the target causal parameter $\eta$ can be estimated using observed data $v$ after obtaining the weights $w_i, i=1,...,n$: $$
\begin{aligned}
	&\operatorname*{argmax}_{\eta}E[l(y_i^*\mid t^*_{max,i},\bar{a}^*_i,\eta)\mid v,\mathcal{J_E}]\\ &=\operatorname*{argmax}_{\eta}\left[\sum\limits_{i=1}^n\pi_iw_il(y_i\mid t_{max,i},\bar{a}_i,\eta)\right]\\
	&=\operatorname*{argmax}_{\eta} \prod\limits_{i=1}^n p(y_i\mid t_{max,i},\bar{a}_i,\eta)^{n\pi_iw_i}\\
	&\equiv q(\eta;v,\theta,\alpha,\pi),
\end{aligned}
$$
Randomly drawing vectors $\pi_{(k)}=(\pi_1^{(k)},...,\pi_n^{(k)}),k = 1,...,l$, and taking $\hat{\eta}_{(k)} := \operatorname*{argmax}_{\eta} q(\eta;$
$v,\theta,\alpha,\pi_{(k)})$ can produce an approximate sample of size $l$ from the posterior distribution of $\eta$ (see \textbf{Web Appendix A} for more details). In our simulation, $n\pi_{(k)}\sim\text{ Multinomial}(n; n^{-1},...,n^{-1})$. Details of the derivation of weights can be found in \textbf{ Web Appendix B}.

\section{SIMULATION}
\label{sec4}
This section introduces a data generation algorithm, followed by the computation of true causal parameters and details of MCMC sampling. The simulation results are then presented. Model and parameter settings of simulation are presented in \textbf{Web Appendix C}. 
The unmeasured confounding $U$ is assumed to follow a Bernoulli distribution (i.e., $U\sim \text{Bernoulli}(\theta_U)$) and the prior of $\theta_U$ is set as an uniform distribution (i.e., $p(\theta_U)=\text{Uniform}(0,1)$). Three scenarios with increasing confounding levels are considered, corresponding to $\delta=0$, $\delta=0.15$, and $\delta=0.3$. 

\subsection{Data generation}
\label{sec4.1}
\textbf{Algorithm} \ref{Algorithm1} is proposed to generate data under $\mathcal{J_O}$ (cf. P{\'e}nichoux, Moreau, and Latouche \cite{penichoux2015simulating}). In Algorithm \ref{Algorithm1}, the initial value of $A(t)$ is set to zero, and the parameter $dt$ controls the discretization level of the interval and, consequently, the approximation level of integrations. The specific value of $dt$ can be determined based on precision requirements. It is assumed that the changes in the states of different processes occur sequentially ($N^{T_{max}}(t)\to L(t)\to A(t)$). 

By substituting the parameter $\theta$ with $\alpha$, and replacing the corresponding conditional models, data under the $\mathcal{J_E}$ can be generated. The key difference between $\mathcal{J_E}$ and $\mathcal{J_O}$ is whether confounding variables affect the generation process of the interventional part.  The model and parameter settings for the non-interventional part under $\mathcal{J_E}$ remain identical to those under $\mathcal{J_O}$. 

\begin{algorithm}[hbpt]
	\caption{Data Generation Algorithm}
	\label{Algorithm1}
	\textbf{Input:} Sample size $n$, precision parameter $dt$, interval parameters $t_R, D_L$, model parameters $\theta$.\\
	\textbf{Output:}  $\{(y_i,t_{max,i}, z_i, u_i, \bar{a}_i, \bar{l}_i), i=1, \dots, n\}$.
	\begin{algorithmic}
		\FOR {$i = 1, \dots, n$}
		\STATE Sample $z_i \sim p(z\mid \theta_Z)$, $u_i \sim p(u\mid \theta_U)$, $l_i(0 - dt) \sim p(l\mid z_i, \theta_L)$ and set $a_i(0 - dt) = 0$  \;
		\STATE Initialize $t = 0, \Delta_{T_{max,i}}(0) = 0$\;
		
		\WHILE {$\Delta_{T_{max,i}}(t) = 0$ and $t < t_R$}
		\IF {$t \in D_L$}
		\STATE  Sample $l_i(t) \sim p(l\mid \mathcal{H}^L_{it}, \theta_L)$\;
		\ELSE
		\STATE Set $l_i(t) = l_i(t - dt)$\;
		\ENDIF
		\STATE Sample $\Delta_{a_i}(t) \sim \text{Bernoulli}(\lambda^A(t\mid \mathcal{H}^A_{it}, \theta_{A1})dt)$\;
		
		\IF{$\Delta_{a_i}(t) = 0$}
		\STATE   Set $a_i(t) = a_i(t - dt)$\;
		\ELSE
		\STATE   Sample $a_i(t) \sim p(a\mid\mathcal{H}^A_{it}, \theta_{A2})$\;
		\ENDIF
		\STATE Set $t = t + dt$\;
		\STATE Sample $\Delta_{T_{max,i}}(t) \sim \text{Bernoulli}(\lambda^{T_{max}}(t\mid\mathcal{H}^{T_{max}}_{it}, \theta_{T_{max}})dt)$\;
		\ENDWHILE
		\STATE Set $t_{max,i} = t$ and sample $y_i \sim p(y\mid t_{max,i}, \bar{a}_i, \bar{l}_i, z_i,u_i, \theta_Y)$\;
		\STATE Save $(y_i, t_{max,i}, z_i, u_i, \bar{a}_i, \bar{l}_i)$\;
		\ENDFOR
	\end{algorithmic}
\end{algorithm}

\subsection{Calculation of true causal parameters}
The target causal parameter $\eta$ can be defined as the parameter of a given regression model $p(y_i\mid t_{max, i},\bar{a}_i,\eta)$ fitted to an infinite sequence of observations from a data generating mechanism characterized by $p(v\mid\mathcal{J_E})$ (cf. Saarela et al. \cite{saarela2015bayesian}). $\eta$  can be approximated up to arbitrary precision by simulation. 

In simulations, we first calculated the posterior means of the marginal parameters of the interventional part, denoted as ($\hat{\alpha}_{T_{max}},\hat{\alpha}_{A}$), through Bayesian posterior inferences. Then, a large number ($50000$ was used in the simulation) of sample points from $\mathcal{J_E}$ were generated. Finally, regression analysis was performed to calculate the true causal parameters. 

The reason that ($\hat{\alpha}_{T_{max}},\hat{\alpha}_{A}$) can be chosen as the marginal parameters of the interventional part to generate data for calculating the true causal parameters is mainly because the true causal parameters do not depend on the marginal parameters (see formula (12) in Saarela et al. \cite{saarela2015bayesian}).

\subsection{Details of MCMC sampling}
Priors and posteriors are obtained through standard procedures. Specifically, independent priors are used, and posterior samples of parameters are acquired by utilizing a No-U-Turn sampler (NUTS) with a target probability distribution proportional to the product of the marginal likelihood and the prior distribution. The probabilistic programming language Stan has a NUTS implementation \cite{carpenter2017stan}, and it is available to R users through the rstan R package \cite{rstan2023}. 

\subsection{Simulation results}
\label{sec4.2}
Five methods are compared in the simulations: (i) the naive unweighted estimator which does not account for confounding ("Naive"); (ii) the frequentist continuous-time IPT weighted estimator with the plug-in estimates $(\hat{\theta}_{T_{max}},\hat{\theta}_{A},\hat{\alpha})$, where the confounder $U$ is ignored in the corresponding models ("CT-IPTW-IgnoreU"); (iii) the frequentist continuous-time IPT weighted estimator with the plug-in estimates $(\hat{\theta}_{T_{max}},\hat{\theta}_{A},\hat{\alpha})$, where the confounder $U$ is ignored in the corresponding models and weights are truncated at the 95th percentile of sample weights ("CT-IPTW-IgnoreU-trunc95"); (iv)  the frequentist continuous-time IPT weighted estimator with the plug-in estimates $(\hat{\theta}_{T_{max}},\hat{\theta}_{A},\hat{\alpha})$, where the confounder $U$ is observed ("CT-IPTW-ObservedU"); (v)  our proposed {\it{Bayesian Continuous-Time MSMs with Unmeasured Confounding}} ("BCT-MSMs-ConsiderU"). 

In practice, $U$ is unobserved and thus the "CT-IPTW-ObservedU" method cannot be applied; it is presented here only as an idealized benchmark for comparison. Considering the advantages of bootstrap variance estimator \cite{austin2016variance} and the complexity of the data structure in this paper, the variances and 95\% confidence intervals for methods (i)-(iv) are computed based on 200 bootstrap samples. For the proposed  "BCT-MSMs-ConsiderU" method, variances and 95\% credible intervals are calculated using 200 posterior estimators.   

Table \ref{Table1} presents the results under three levels of confounding ($\delta=0,0.15,0.3$) over $300$ simulation rounds for various estimators. 

First, when $\delta = 0$, there is no confounding, and thus the "Naive" method performs well in this scenario. Second, "CT-IPTW-ObservedU" can provide a more accurate estimate and a smaller Monte Carlo standard deviation of the point estimates (SD) than "CT-IPTW-IgnoreU" as expected. Third, "CT-IPTW-IgnoreU" suffers from model misspecification due to neglecting the variable U, resulting in extreme weights and leading to abnormally large bias and SD of the estimator. Although "CT-IPTW-IgnoreU-trunc95" significantly alleviates this issue, it still performs worse compared to "CT-IPTW-ObservedU" and "BCT-MSMs-ConsiderU". Fourth, the proposed method, "BCT-MSMs-ConsiderU", provides approximately unbiased estimates of the target causal parameters and demonstrates comparable 95\% CP (confidence/credible interval coverage probability) and LCI (average length of 95\% confidence/credible interval) to the "CT-IPTW-ObservedU" method.

Overall, the proposed method outperforms all other practically feasible approaches and demonstrates performance comparable to the idealized benchmark method ("CT-IPTW-ObservedU").

\begin{table}[htbp]
	\centering
	\caption{Simulation results. $n$ $(=400)$ sample points and $300$ replications. The columns correspond to the estimator, bias, Monte Carlo standard deviation of the point estimates (SD), mean standard error estimate (SE), 95\% confidence/credible interval coverage probability (CP), and average length of 95\% confidence/credible interval (LCI).}
	\renewcommand\arraystretch{1.5} 
	\begin{tabular}{cccccccc}
		\hline
		\textbf{Scenario}& \multicolumn{2}{c}{\textbf{Estimator}}           & \textbf{Bias} & \textbf{SD} & \textbf{SE} & \textbf{95\% CP} & \textbf{LCI} \\ \hline
		\multirow{9}{*}{\begin{tabular}[c]{@{}c@{}}$\delta=0$\\ $\eta_2=2.051$,\\ $\eta_3=2.066$\end{tabular}}    & \multirow{2}{*}{Naive}                & $\eta_2$         & -0.023        & 0.211       & 0.213       & 94.932           & 0.814        \\
		&                                       & $\eta_3$          & -0.042         & 0.282       & 0.285       & 94.932           & 1.098        \\ \cline{2-8} 
		& \multirow{2}{*}{CT-IPTW-IgnoreU}    & $\eta_2$         & 5.265         & 19.907      & 37.883       & 99.324           & 46.537        \\
		&                                       & $\eta_3$          & 11.243         & 77.824       & 84.634       & 98.986           & 52.397        \\ \cline{2-8} 
		& \multirow{2}{*}{CT-IPTW-IgnoreU-trunc95}    & $\eta_2$         & 0.806         & 1.331      & 4.076       & 95.946           & 6.789       \\
		&                                       & $\eta_3$          & 0.597         & 1.107       & 2.549       & 95.270           & 6.450        \\
		\cline{2-8} 
		& \multirow{2}{*}{CT-IPTW-ObservedU}   & $\eta_2$         & -0.026         & 0.212       & 0.214       & 94.595           & 0.819        \\
		&                                       & $\eta_3$          & -0.046         & 0.284       & 0.287       & 95.270           & 1.105        \\ \cline{2-8} 
		& \multirow{2}{*}{BCT-MSMs-ConsiderU} & $\eta_2$          & \bf{-0.020}        & \bf{0.235}       & \bf{0.224}       & \bf{93.581}           & \bf{0.859}        \\
		&                                       & $\eta_3$      & \bf{-0.041}         & \bf{0.316}       & \bf{0.301}       & \bf{91.892}           & \bf{1.157}        \\ \hline
		\multirow{9}{*}{\begin{tabular}[c]{@{}c@{}}$\delta=0.15$\\ $\eta_2=2.188$,\\ $\eta_3=1.939$\end{tabular}}  & \multirow{2}{*}{Naive}                & $\eta_2$          & 0.266        & 0.184       & 0.171       & 62.000          & 0.656        \\
		&                                       & $\eta_3$          & 0.334         & 0.219       & 0.202      & 58.667           & 0.776        \\ \cline{2-8} 
		& \multirow{2}{*}{CT-IPTW-IgnoreU}     & $\eta_2$          & -658.130        & 22724.833       & 25983.171       & 98.667           & 635.929        \\
		&                                       & $\eta_3$         & 1251.950         & 17091.739       & 13717.461       & 97.667          & 1313.433        \\ \cline{2-8} 
		& \multirow{2}{*}{CT-IPTW-IgnoreU-trunc95}     & $\eta_2$          & 1.185        & 2.172       & 7.340       & 89.333           & 7.624        \\
		&                                       & $\eta_3$         & 0.693         & 0.870       & 2.975       & 89.000          & 4.871        \\ \cline{2-8} 
		& \multirow{2}{*}{CT-IPTW-ObservedU}   & $\eta_2$          & 0.098         & 0.238       & 0.217       & 89.667           & 0.831        \\
		&                                       & $\eta_3$         & 0.126         & 0.296       & 0.267       & 89.000           & 1.023        \\  \cline{2-8} 
		& \multirow{2}{*}{BCT-MSMs-ConsiderU} & $\eta_2$          & \bf{0.027}        & \bf{0.235}       & \bf{0.214}       & \bf{92.667}           & \bf{0.808}        \\
		&                                       & $\eta_3$          & \bf{0.032}         & \bf{0.299}       & \bf{0.267}       & \bf{92.333}           & \bf{1.009}        \\  \hline
		\multirow{9}{*}{\begin{tabular}[c]{@{}c@{}}$\delta=0.3$\\ $\eta_2=2.389$,\\ $\eta_3=1.962$\end{tabular}} & \multirow{2}{*}{Naive}                & $\eta_2$          & 0.383        & 0.166       & 0.170       & 31.333            &0.652        \\
		&                                       & $\eta_3$         & 0.480         & 0.181       & 0.187       & 21.000           & 0.720        \\ \cline{2-8} 
		& \multirow{2}{*}{CT-IPTW-IgnoreU}     & $\eta_2$          & 47102.422        & 500698.213      & 714157.055       & 99.667           & 21240.271       \\
		&                                       & $\eta_3$         & 387999.735         & 6338439.854       & 5302224.045      & 98.000           & 41154.657       \\\cline{2-8} 
		& \multirow{2}{*}{CT-IPTW-IgnoreU-trunc95}     & $\eta_2$          & 0.565       & 0.944     & 3.361       & 91.000           & 4.280       \\
		&                                       & $\eta_3$         & 0.542         & 0.531      & 1.734      & 82.333          & 3.440       \\  \cline{2-8} 
		& \multirow{2}{*}{CT-IPTW-ObservedU}   & $\eta_2$          & 0.183        & 0.371       & 0.337       & 85.667           & 1.214        \\
		&                                       & $\eta_3$          & 0.223         & 0.393       & 0.362       & 83.667           & 1.337        \\  \cline{2-8} 
		& \multirow{2}{*}{BCT-MSMs-ConsiderU} & $\eta_2$         & \bf{0.054}        & \bf{0.321}      & \bf{0.373}       &\bf{92.333}           & \bf{1.077}        \\
		&                                       & $\eta_3$          & \bf{0.057}         & \bf{0.367}       & \bf{0.361}       & \bf{94.000}           & \bf{1.261}        \\  \hline
	\end{tabular}
	\label{Table1}
\end{table}

\section{ANALYSIS OF THE CAUSAL RELATIONSHIP BETWEEN OXYTOCIN ADMINISTRATION PROCESS AND POSTPARTUM HEMORRHAGE}
\label{sec5}
\subsection{Study background}
\label{sec5.1}
The literature has found increased PPH cases over the past few decades \cite{lutomski2012increasing}. The use of oxytocin during labor has also been on the rise \cite{buchanan2012trends}, leading to concerns about a potential link between oxytocin use and an elevated risk of PPH. It is doubted that prolonged oxytocin use during labor may deplete oxytocin receptors in the myometrium, potentially increasing the chances of uterine atony, retained placenta, and PPH \cite{bateman2010epidemiology,endler2012epidemiology,robinson2003oxytocin}.

The administration of oxytocin occurs continuously and dynamically within a closed interval, forming a complex continuous-time dynamic treatment (Figure \ref{fig:Fig1}). Controlling for confounders in this scenario is even more challenging than in discrete-time settings. Conducting a randomized controlled trial is not feasible or ethical because oxytocin is the primary treatment for dystocia; thus, analyzing real-world data may be the only viable option to address this important clinical question. 

\subsection{Data source}
The Consortium on Safe Labor (CSL) is a retrospective observational study including 19 hospitals from 12 clinical centers across 9 American College of Obstetricians and Gynecologists US districts \cite{zhang2010contemporary}. Data are extracted from electronic medical records on maternal demographic characteristics, medical history, reproductive and prenatal history, labor and delivery summary, and postpartum information. Detailed information on CSL is provided in Zhang et al. \cite{zhang2010contemporary}.

\subsection{Analysis}
\label{sec5.2}


Based on the dataset used in Zhu et al. \cite{zhu2024oxytocin}, the complete cases consisting of 6324 subjects are selected in this paper.  Similar criteria as in Zhu et al. \cite{zhu2024oxytocin} are used to select variables into the analysis; one continuous-time dynamic treatment ($A(t)=$ Oxytocin dosage (mu/min) at time $t$), one outcome variable ($Y=$ Estimated blood loss (100 ml)), one discrete-time varying covariate ($L(t)=$ The degree of cervical dilation at time $t$), five variables representing different time, twelve discrete covariates, twelve continuous covariates are included. The names of specific variables are listed in \textbf{Web Appendix D}. In addition, a variable transformation ($\tilde{x}=\log(x+1)$) is applied to treatment $A(t)$ to facilitate easier modeling, and variable standardization (scaling) is performed on continuous-type covariates. $D_L$ is taken as the union of the observation times of $L(t)$. For individual $i$, assume $\bar{l}_i$ is measured at $t^L_{ij}, \, j = 0, \dots, J_i$. Although the measurement times of $L(t)$ may differ across individuals (i.e., $J_i$ varies), only the observed $l_i(t)$ impacts $a_i(t)$ and our method does not require modeling the distribution of $\bar{L}$. Thus, we set $l_i(t)$ to $l_i(t^L_{ik})$ for $t \in D_L \cap (t^L_{ik}, t^L_{i,k+1}), k=0,...,J_i-1$.

This study aims to evaluate the causal effects of the oxytocin administration process. The potential outcome model is set as the model (\ref{equa_3_1}), and other model settings are the same as Section \ref{sec4} (see \textbf{Web Appendix C}). In clinical practice, to reduce the risk of postpartum hemorrhage, active management of the third stage of labor is commonly performed \cite{hersh2024third}. The active management of the third stage of labor involves the administration of prophylactic uterotonic drugs just before or immediately after delivery \cite{gizzo2013uterotonic}. Therefore, it is reasonable to assume that the effect of oxytocin is stronger closer to the delivery time, which clinically supports the validity of the model (\ref{equa_3_1}).

Three approaches ("Naive"; "CT-IPTW-IgnoreU-trunc95"; "BCT-MSMs-ConsiderU") discussed in Section \ref{sec4} were implemented and results were organized in Table \ref{Table2}. For "BCT-MSMs-ConsiderU", weights were truncated at the 98th percentile to ensure no extreme weights. The results of "Naive" and "CT-IPTW-IgnoreU-trunc95" were based on 500 bootstrap samples, while those of "BCT-MSMs-ConsiderU" were based on 500 posterior samples. 

Table \ref{Table2} shows the results of the three methods. Firstly, all three methods yield statistically significant estimates of $\eta_2$ (greater than zero). Secondly, "BCT-MSMs-ConsiderU" produces a higher estimate of $\eta_2$ and lower estimate of $\eta_3$ ($\hat{\eta}_2=0.067, \hat{\eta}_3=0.959$) compared to the "Naive" ($\hat{\eta}_2=0.032, \hat{\eta}_3=1.051$) and "CT-IPTW-IgnoreU-trunc95" ($\hat{\eta}_2=0.041, \hat{\eta}_3=3.121$) methods. For "BCT-MSMs-ConsiderU", the 95\% credible intervals of $\eta_2$ and $\eta_3$  in Table \ref{Table2} are $[0.035,0.135]$ and $[0.000,4.051]$, respectively. This indicates that, from a clinical perspective, an increase in the duration and dose of oxytocin may lead to a slight increase in postpartum bleeding. For example, if an individual begins oxytocin administration at the $4$th hour after admission and maintains \( A(t) = 3 \) 
until delivery (which occurs at the $12$th hour after admission), the cumulative effect of this regimen is (approximately) at least $\hat{\eta}_{2,low} \int_4^{12} A(t) \exp \left( -\hat{\eta}_{3,up}|t-12|/12 \right) dt \approx 0.29$, corresponding to an increase of \( 0.29 \times 100 = 29 \, \text{ml} \) in postpartum hemorrhage volume, where $\hat{\eta}_{2,\text{low}}=0.035$ and $\hat{\eta}_{3,\text{up}}=4.051$ are the lower and upper bounds of the credible intervals of the proposed method, respectively. 

In summary, although the estimated parameters and the example provided suggest that oxytocin use may lead to an increase in postpartum blood loss, the increase is relatively mild. Clinically, it cannot be concluded that oxytocin significantly increases the risk of postpartum hemorrhage, and further evidence may be needed for a more thorough analysis. 
\begin{table}[htbp]
	\centering
	\caption{Real data analysis results. The columns correspond to the estimator, mean point estimate, Monte Carlo standard deviation of the point estimates (SD), and 95\% confidence/credible interval. Results are based on 500 bootstrap/posterior samples.}
	\renewcommand\arraystretch{1.5} 
	\begin{tabular}{cccccc}
		\hline
		&\multicolumn{2}{c}{\textbf{Estimator}}& \textbf{Mean} & \textbf{SD} & \textbf{95\% CI}    \\ 
		\hline
		&\multirow{2}{*}{Naive}                & $\eta_2$ & 0.032         & 0.010       & [0.019,0.057]       \\
		&                                       & $\eta_3$ & 1.051         & 0.934      & [0.000,3.309]     \\ 
		\hline
		& \multirow{2}{*}{CT-IPTW-IgnoreU-trunc95}    & $\eta_2$ & 0.041         & 0.125        & [0.010,0.134]    \\
		&                                       & $\eta_3$ & 3.121         & 32.496       & [0.000,6.928]      \\ \hline
		& \multirow{2}{*}{BCT-MSMs-ConsiderU} & $\eta_2$ & \bf{0.067}        & \bf{0.030}      & \bf{[0.035,0.135]}    \\
		&                                       & $\eta_3$ & \bf{0.959}         & \bf{1.281}      & \bf{[0.000,4.051]} \\
		\hline
	\end{tabular}
	\label{Table2}
\end{table}

\section{DISCUSSION}
\label{sec6}
Motivated by inconclusive real-world evidence for the impact of the oxytocin administration process on postpartum hemorrhage, a Bayesian estimation approach for continuous-time marginal structural models has been developed. This approach addresses the challenge of handling unmeasured confounding when making causal inferences about continuous-time dynamic treatments. Unmeasured confounders $U$ are marginalized in the likelihoods and considered in estimating weights. In model (\ref{equa_3_1}), the $t_{max}$ in the numerator of $\exp\left( -\eta_3 |t-t_{\text{max}}|/t_{max} \right)$ can be replaced with other meaningful time points depending on the specific context.

The proposed method is easy to extend to the settings involving latent individual level "frailty" variables as Saarela et al. \cite{saarela2015bayesian}, and discrete-time varying unmeasured confounding. Individual level "frailty" variables are determinants of both the outcome and the intermediate variables, which relates to the "null paradox" discussed by Robins and Wasserman \cite{robins2013estimation}. If $U$ represents latent individual level "frailty" variables, our method only requires modifications to $p(v\mid \mathcal{J_O})$ and $p(v\mid \mathcal{J_E})$, and subsequent derivations can be obtained similarly. 

The use of parametric models is primarily due to their simplicity and interpretability. However, given their limitations, future work could explore extending our method to nonparametric models. Additionally, our approach could be enhanced by improving numerical computational efficiency, such as by approximating the integral of intensity functions as suggested by Deutsch and Ross \cite{deutsch2022estimating}.

\bibliographystyle{unsrt}  
\bibliography{wileyNJD-AMA}

\end{document}